\documentclass[12pt]{article}
\usepackage{graphicx}
\input epsf
\usepackage{amssymb}
\parskip=\medskipamount
\def\ID{\relax{\rm l\kern-.18 em D}}
\def\IE{\relax{\rm l\kern-.18 em E}}
\def\IK{\relax{\rm l\kern-.18 em K}}
\def\IL{\relax{\rm I\kern-.18 em L}}
\def\IN{\relax{\rm I\kern-.18 em N}}
\def\IR{\relax{\rm I\kern-.18 em R}}
      
\def\uno{\relax{\rm 1\kern-.18 em l}}

\def\IK{\relax{\rm l\kern-.18 em K}}
\def\IL{\relax{\rm I\kern-.18 em L}}
\def\IN{{\Bbb N}}
\def\IR{{\Bbb  R}}

\def\pd#1#2{\frac{\partial #1}{\partial#2}}

\def\frac#1#2{{#1\over #2}}

\def\ptos{\leaders\hbox to 2mm{\hfil{.}\hfil}\hfill}
\def\\{\hfill\break}

\def\<#1>{\langle#1\rangle}

\font\tenfrak=eufm10  \font\sevenfrak=eufm7  \font\fivefrak=eufm5
\newfam\frakfam
\textfont\frakfam=\tenfrak\scriptfont\frakfam=\sevenfrak
\scriptscriptfont\frakfam=\fivefrak

\font\tengoth=eufm10 scaled\magstep1 \font\sevengoth=eufm7
\font\fivegoth=eufm5
\newfam\gothfam
\textfont\gothfam=\tengoth\scriptfont\gothfam=\sevengoth
   \scriptscriptfont\gothfam=\fivegoth

\newtheorem{proposicion}{Proposition}

\begin{document}

\title{Higher-order superintegrability of a \\ Holt related potential }

\author{
R.  Campoamor-Stursberg$\ddagger\,^{a)}$,
J.F. Cari\~nena$\dagger\,^{b)}$, and
M.F. Ra\~nada$\dagger\,^{c)}$ \\
$\dagger$
   {\sl Departamento de F\'{\i}sica Te\'orica and IUMA, Facultad de Ciencias} \\
   {\sl Universidad de Zaragoza, 50009 Zaragoza, Spain}  \\
$\ddagger$
   {\sl Departamento de Geometr\'{\i}a y Topolog\'{\i}a and IMI} \\
   {\sl Universidad Complutense, 28040 Madrid, Spain}
}
\date{Thu, 26 Sep 2013  }
\maketitle

\begin{abstract}  
In a recent paper, Post  and Winternitz studied  the  properties of two-dimensional Euclidean  potentials that are linear in one of the two Cartesian variables. In particular,  they proved  the existence of a potential endowed with an integral of third-order and an integral of fourth-order. 
In this paper we show that these results can be obtained in a  more simple and direct way by noting that this potential is directly related with the Holt potential. It is proved that the existence of a potential with higher order superintegrability is a direct consequence of the integrability of the family of Holt type potentials.

\end{abstract}

\begin{quote}
Higher-order  constants of motion. 

{\sl Running title:}{\enskip}
Superintegrability of a Holt related potential.


\end{quote}

\vfill
\footnoterule
{\noindent\small
$^{a)}${\it E-mail address:} {rutwig@ucm.es } \\
$^{b)}${\it E-mail address:} {jfc@unizar.es }  \\
$^{c)}${\it E-mail address:} {mfran@unizar.es }
\newpage

\section{Introduction}

It is well known \cite{Perelomov} that there are two different classes of integrable systems :   (i)  separable  and (ii) nonseparable.
\begin{itemize}
\item[(i)] Systems that admit  Hamilton-Jacobi (Schr\"odinger) separability in a particular coordinate system  are integrable  with constants of the motion  (the first one is the Hamiltonian itself) which are quadratic in the momenta (if the constant is determined by an exact Noether symmetry then it is linear).  For example, in the Euclidean plane, if the potential is separable in Cartesian coordinates then the Hamiltonian is integrable with the two  one-dimensional energies as constants of the motion. 
Two  other particular cases are 

\begin{itemize}
 
\item Polar separability: If the potential can be written as $V=A(r) + B(\phi)/r^2$ then the system is integrable with the following second constant of the motion
$$
 K_2^{(2)}  = p_\phi^2 + 2 B(\phi) \,.
 $$

\item  Parabolic separability: If the potential can be written as $V=(A(\alpha) + B(\beta))/(\alpha^2 + \beta^2)$ then the Hamiltonian is integrable with a second constant of the motion of Runge-Lenz type 
$$
 K_2^{(2)} =  p_\alpha^2 + 2 A(\alpha) - 2 H \alpha^2 =  \frac{\alpha^2p_\beta^2 - \beta^2 p_\alpha^2}{\alpha^2 + \beta^2} + 2\,\frac{\alpha^2B(\beta) - \beta^2 A(\alpha)}{\alpha^2 + \beta^2} \,. 
$$

\end{itemize}
\item[(ii)] Nonseparable systems. They are integrable but with higher-order  constants of motion (higher-order means higher order than two in the momenta). Very few systems of this class are known. We mention the Holt   \cite{Ho82}  and the Fokas-Lagerstrom \cite{FoL80} potentials that are both endowed with a cubic integral (see \cite{GrDR84}--\cite{CpSt13} for some other related papers). 
 \end{itemize}

If we consider not just integrability but superintegrability then we have three possibilities: (1)  Superseparable, (2) Separable, and (3) Nonseparable. Most known superintegrable systems are of type (1); concerning type (2), that are endowed with a mix of both quadratic and higher order constants, we mention two recently studied systems: the TTW system (related to the harmonic oscillator) \cite{TTW09}-\cite{Ra12b} and the PW system (related to the Kepler problem) \cite{PostWint10}-\cite{Ra13}. The study of type (3) systems, with all the constants of higher order (up to the Hamiltonian), began very recently and, in fact, the discovery of one of these systems must be considered as a very important result. 

In a recent paper, Post and  Winternitz \cite{PostW11}, making use of some previous results  obtained in \cite{GrW02}, studied the existence, in the  real Euclidean plane, of  potentials linear in one of the two Cartesian variables that admit  third-order or  fourth-order  integrals (the  study was made for quantum systems but the results are also valid for the corresponding classical systems by removing some terms proportional to $\hbar^2$). They derive, using some rather complex and elaborate mathematics (a direct calculus involves a great number of coupled equations), a nonseparable quantum superintegrable system,  endowed with an integral of third-order and an integral of fourth-order. This result is certainly important and suggests the convenience of studying the properties of this particular potential  (a study of the Lie symmetries was recently presented in \cite{NucP12}).  The main purpose of this paper is to show that this superintegrable potential (as well as the two integrals)  can be obtained in a direct and rather simpler fashion, making use of the properties of the Holt potential. 
 
 The paper  is organized as follows. In section 2 we recall the main properties of the Holt potentials and then in section 3 we prove the superintegrabilty and we obtain the explicit expressions of the two integrals. Finally in section 4 we make some comments and we present some  open questions.

\section{Potentials of  Holt }

In 1982 Holt  studied  \cite{Ho82}  the existence of integrable Hamiltonians in two degrees of freedom using a generalization of the Whittaker method as an approach (the original Whittaker method was mainly related to quadratic integrals) and proved that  following potential
\begin{equation}
  V_{h1}   = \frac{4 x^2 + 3 y^2}{y^{ 2/3}}
\end{equation}
was endowed with the following cubic integral 
$$
  J_{h1}^{(3)}  = 2 p_x^3 + 3 p_x p_y^2 + 12 \Bigl( \frac{2 x^2 - 3 y^2}{y^{2/3}} p_x + 6 x y^{1/3} p_y \Bigr)\,.
$$
This means that $V_{h1}$ belongs to the family (ii) of integrable but not separable potentials.

It has been proved  the existence, in addition to the original potential $ V_{h1}$, of  two other  integrable potentials of Holt type  (see e.g.  \cite{Ts99}  or \cite{CpSt13}). They are : 
\begin{itemize}
\item[(h2)]  The following potential 
\begin{equation}
  V_{h2}   =  \frac{2 x^2 + 9 y^2}{y^{ 2/3}} \,,
\end{equation}
which is nonseparable, and is also integrable with a constant of the motion of fourth order in the momenta
$$
  J_{h2}^{(4)}  =  p_x^4 + 2 p_x^2 p_y^2 + 8 \Bigl(x^2 y^{-2/3} p_x^2 +   6\,  x y^{1/3} p_x p_y +   36  \, x^2 y^{2/3} \Bigr)\,.
$$
\item[(h3)]  The following potential 
\begin{equation}
  V_{h3}   = \frac{x^2 + 12 y^2}{y^{ 2/3}} 
\end{equation}
which is also nonseparable but integrable with a constant of the motion of sixth order in the momenta  given by 
$$
  J_{h3}^{(6)}  =  p_x^6 + 3 p_x^4 p_y^2 +   6  \Bigr( x^2/y^{2/3} + 3 y^{4/3}  \Bigr)\, p_x^4 + 72 x  y^{1/3}  p_x^3 p_y +   648\, x^2  y^{2/3} p_x^2  +  648\,  x^4 \,. 
$$
\end{itemize}

\section{Superintegrability of a Holt related potential }

An important property is that potential $V_{h1}$, which has been introduced as a single potential,  admits a generalization.  In fact, it can  considered as the first term in a more general potential given by the following linear combination 
\begin{equation}
  V_{h1}(k)   =   k_1\, \Bigl(\frac{4 x^2 + 3 y^2}{y^{ 2/3}}\Bigr) \,+\, k_2\,  \frac{x}{y^{ 2/3}} \,+\,   \frac{k_3}{y^{ 2/3}} 
\end{equation}
where $k_i$, $i=1,2,3$, are arbitrary constants.  In this more general case the third-order integral is given by  
$$
  J_{h1}^{(3)}(k)  = 2 p_x^3 + 3 p_x p_y^2 + 12 k_1\Bigl( \frac{2 x^2 - 3 y^2}{y^{2/3}} p_x + 6 \,x y^{1/3} p_y \Bigr) + k_2 \Bigl(\frac{6 \,x}{y^{2/3}} p_x + 9 \,y^{1/3} p_y  \Bigr) + \frac{ 6 \,k_3}{ y^{2/3}} p_x  \,.
$$
The same is true for the second Holt potential. That is, the potential $V_{h2}$ can  also be considered as the first term of a more general potential given by the following linear combination 
\begin{equation}
  V_{h2}(k)   =   k_1\, \Bigl( \frac{2 x^2 + 9 y^2}{y^{ 2/3}}\Bigr) \,+\, k_2\,  \frac{x}{y^{ 2/3}} \,+\,   \frac{k_3}{y^{ 2/3}} 
\end{equation}
where $k_i$, $i=1,2,3$, are arbitrary constants. The integral of motion, 
that is of fourth order in the momenta, is given by 
$$
  J_{h2}^{(4)}(k)  = p_x^4 + 2 p_x^2 p_y^2 + \frac{4 (2 k_1 x^2 + k_2 x + k_3)}{y^{2/3}} p_x^2 +  12 (4 k_1 x + k_2) y^{1/3} p_x p_y + 18 (4 k_1 x + k_2)^2 y^{2/3} \,.
$$

The important point is that although $V_{h1}(k)$ and $V_{h2}(k)$ are different potentials (both integrable but each one in a different way) they have however the $k_2$ and $k_3$ dependent terms in common. So, if we denote by $U$ the two-parameter dependent family of potentials 
\begin{equation}
 U = \lim_{k_1\to 0} V_{h1}(k) = \lim_{k_1\to 0} V_{h2}(k) = k_2\,  \frac{x}{y^{ 2/3}} \,+\,   \frac{k_3}{y^{ 2/3}} 
\end{equation}
then this family  of potentials possesses the following two constants of motion 
\begin{eqnarray}    
  K_2^{(3)}  &=&  2 p_x^3 + 3 p_x p_y^2 +  k_2 \Bigl(\frac{6 \,x}{y^{2/3}} p_x + 9 \,y^{1/3} p_y  \Bigr) + \frac{ 6 \,k_3}{ y^{2/3}}\, p_x  \label{K2}\\
  K_3^{(4)}  &=&  p_x^4 + 2 p_x^2 p_y^2 + \frac{4 (k_2 x + k_3)}{y^{2/3}} p_x^2 +  12  k_2 y^{1/3} p_x p_y + 18 k_2^2\, y^{2/3} \label{K3}
\end{eqnarray}

As the potential $U$ is part of two different families of potentials that are integrable in a different way, and therefore with different constants of motion, we can conclude that the system is superintegrable. 
The following proposition summarizes the situation. 
 
\begin{proposicion} The potential
$$
   U =  k_2\,  \frac{x}{y^{ 2/3}} \,+\,   \frac{k_3}{y^{ 2/3}} 
$$
is not separable but it is  superintegrable with two integrals of motion that are of the third and fourth order in the momenta, respectively, given by (\ref{K2}) and (\ref{K3}).
\end{proposicion}

Consequently, the system is superintegrable, the three constants of motion being $H, K_2^{(3)}$ and $ K_3^{(4)}$, and we recover in a much simpler way the result of \cite{PostW11}.

\subsection{The family of the third Holt potential}

We can also consider the three-parameter dependent potential generated by $V_{h3}$
\begin{equation}
  V_{h3}(k)   =   k_1\, \Bigl( \frac{x^2 + 12 y^2}{y^{ 2/3}}  \Bigr) \,+\, k_2\,  \frac{x}{y^{ 2/3}} \,+\,   \frac{k_3}{y^{ 2/3}} 
\end{equation}
where $k_i$, $i=1,2,3$, are arbitrary constants. The integral of motion, 
that now is of sixth order in the momenta, is given by 
$$
  J_{h3}^{(6)}(k)  = p_x^6 + 3 p_x^4 p_y^2 + J_{40} p_x^4 + J_{31} p_x^3 p_y + J_{20} p_x^2 + J_0 \,, 
$$
where $J_{40}$, $J_{31}$, $J_{20}$, and $J_{0}$, are given by 
\begin{eqnarray*}  
J_{40}  &=&  6  \,(k_1 x^2 + 3 k_1 y^2 + k_2  x + k_3)y^{-2/3} \cr
J_{31}  &=&  36 (2 k_1 x + k_2) y^{1/3}  \cr 
J_{20}  &=&  162 (2 k_1 x + k_2)^2 y^{2/3}  \cr 
J_0 &=& 324 x ( k_1 x + k_2) (2 k_1^2 x^2 + 2 k_1 k_2 x + k_2^2 )
\end{eqnarray*}  
Now, if we consider the limit $k_1\to 0$ we obtain a new integral of motion for the potential $U$
$$
  K_4^{(6)}  =p_x^6 + 3 p_x^4 p_y^2  
  + 6\Bigl(\frac{k_3 + k_2 x}{y^{2/3}} \Bigr)p_x^4+   36 k_2 y^{1/3} p_x^3 p_y  
  + 162 k_2^2   y^{2/3} p_x^2 + 324 k_2^3 x \,. 
$$
Of course this new function is not independent of the three fundamental functions (the Hamiltonian $H$, $K_2^{(3)}$ and $K_3^{(3)}$). We have verified that the relation is the following 
$$
  K_4^{(6)}  = 18  H K_3^{(4)} - 2(K_2^{(3)})^2 - 324  k_2^2 k_3  \,. 
$$

\subsection{Poisson brackets and Hamiltonian vector fields}

The Hamiltonian vector fields $X_2$ and $X_3$ corresponding to $ K_2^{(3)}$ and $ K_3^{(4)}$ generate infinitesimal symmetries of the dynamics and one could expect that the commutator of both to provide a new infinitesimal symmetry. However, as proved in \cite{PostW11},  both integrals of motion are such that  
\begin{equation}
  \Bigl\{K_3^{(4)}\, , K_2^{(3)}\Bigr\} =  108 \,k_2^3 \,. 
\end{equation}
Therefore, if $\omega$ is the natural symplectic form in the phase space, using that 
\begin{equation}
  i\bigl([X_2 , X_3]\bigr)\,\omega = -\,d \Bigl\{K_2^{(3)},K_3^{(4)}\Bigr\} ,
\end{equation}
we find the both vector fields commute, that is $[X_2 , X_3]=0$. 
 In other words, there is a symplectic action 
of the Abelian Lie algebra $\mathbb{R}^2$ but the comomentum map is not an homomorphism. This happens for symplectic action of Lie groups whenever the momentum map is not equivariant.

We can also consider the other two Poisson brackets, given by
$$
  \Bigl\{K_4^{(6)}\, , K_2^{(3)}\Bigr\} =  1944\, k_2^3\, H 
 {\quad}{\rm and}{\quad}
  \Bigl\{K_4^{(6)}\, , K_3^{(4)}\Bigr\} =  432\, k_2^3\, K_2^{(3)} \,. 
$$
and the corresponding commutation relations for the associated Hamiltonian vector fields are:
$$ 
 [X_2, X_4] = 1944\, k_2^3\,\Gamma_H, \qquad 
 [X_3, X_4] = 432\, k_2^3 \,X_2.
$$ 
where $\Gamma_H$ is the dynamical vector field
$$ 
  \Gamma_H = p_x\pd{}{x} + p_y\pd{}y  - \frac{k_2}{y^{ 2/3}}\pd{}{p_x} 
  + \frac{2}{3} \frac{k_2\,x+k_3}{y^{ 5/3}} \pd{}{p_y} .
$$

The integrals of motion are a function group \cite{Eisenhart} and we can take different choices for bases.
In the first choice, as pointed out in \cite{PostW11}, the chosen  integrals of motion generate a finite dimensional decomposable Lie algebra that is a sum of a Heisenberg Lie algebra spanned by $ K_2^{(3)},  K_3^{(4)}, I$ and the one-dimensional Lie algebra generated by $H$. 
In the second choice, the basis functions  $K_2^{(3)},  K_4^{(6)}, H$ close a Heisenberg Lie algebra, with center $H$.

\section{Final comments  } 

We have proved that the higher-order superintegrability of the potential $U$ can be considered as a consequence of the properties of the Holt type potentials. In this way we have proved the superintegrability (an we have obtained the integrals of motion) in a very simple and elegant way.

In fact, the Holt potentials have been studied by many authors making use of different approaches (Painlev\'e analysis \cite{GrDR84},  direct method \cite{Hiet87},  Lax equations \cite{Ts99}, relation with the Drach potentials \cite{RaS01}, \cite{Ts00}, Lie symmetries \cite{DaSop99}, scaling symmetries \cite{CpSt13}), but in spite of this, we see that they are endowed with certain properties that still remain to be studied. 
Now, concerning the relation of the superintegrable potential $U$ with the Holt potentials we consider that there two interesting points to be studied:
(i) the existence of higher-dimensional versions of the Holt potentials  \cite{GrDRH85} as a method for obtaining potentials similar to $U$ but with three or more degrees of freedom, and (ii) the duality and coupling-constant metamorphosis  between integrable Hamiltonian systems  \cite{HieGDR84}, \cite{SerBl08}, and,  more specifically,  the relation of the Holt potential with the H\'enon-Heiles-type Hamiltonians  \cite{Hiet83} can also lead to the construction of other superintegrable systems.

\section*{Acknowledgments}

This work was supported by the research projects MTM--2012--33575 (MICINN, Madrid)  and DGA-E24/1 (DGA, Zaragoza) and MTM2010-18556  (MICINN, Madrid).

{\small
 }
\end{document}